\documentclass[%
aps,nofootinbib,
notitlepage,
superscriptaddress,
twocolumn,
nobibnotes,
 amsmath,
 amssymb,
 prd,
]{revtex4-2}

\usepackage{graphicx}
\usepackage{dcolumn}
\usepackage{bm}
\usepackage{graphicx}
\usepackage{xcolor}
\usepackage{hyperref}

\definecolor{Maroon}{cmyk}{0, 0.87, 0.68, 0.32}
\usepackage{orcidlink}

\hypersetup{
    colorlinks = true,
    linkcolor  = Maroon,
    urlcolor   = Maroon,
    citecolor  = Maroon,
    pdfborder  = {0 0 0}
}

\begin{document}

\title{Measuring neutrino mass and asymmetry through galaxy pairwise peculiar velocity}

\author{Wangzheng Zhang\,\orcidlink{0000-0003-0102-1543}}
 \email{zhang@iap.fr}
\affiliation{
 Institut d'Astrophysique de Paris, UMR~7095, CNRS, Sorbonne Universit\'e, 98~bis~boulevard Arago, 75014~Paris, France
}
\affiliation{
 Department of Physics, the Chinese University of Hong Kong, ST, NT, Hong Kong
}

\author{Ming-chung Chu\,\orcidlink{0000-0002-1971-0403}}%
\affiliation{
 Department of Physics, the Chinese University of Hong Kong, ST, NT, Hong Kong
}

\author{Shihong Liao\,\orcidlink{0000-0001-7075-6098}}%
\affiliation{
Key Laboratory for Computational Astrophysics, National Astronomical Observatories, Chinese Academy of Sciences, Beijing 100101, China
}
\affiliation{
School of Astronomy and Space Science, University of Chinese Academy of Sciences, Beijing 100049, China
}

\date{\today}

\begin{abstract}
Cosmic neutrinos are among the most abundant fermions in the Universe, yet the values of their masses and chemical potentials remain uncertain. In this Letter, we present the first constraints on the total neutrino mass $M_\nu$ and the neutrino asymmetry parameter $\eta^2$ derived from the mean galaxy pairwise peculiar velocity in the quasi-linear and nonlinear regimes. We develop a simulation-based analysis pipeline that connects neutrino properties to predictions of galaxy pairwise velocity, and apply it to galaxy data from the Cosmicflows-4 grouped catalog.
Our analysis is performed within two independent cosmological frameworks, based on cosmological parameters derived from Cosmic microwave background (CMB) and local distance ladder measurements, respectively. By performing fits to the galaxy pairwise velocity, we obtain consistent constraints from both frameworks. Quoting posterior means with 68\% CL, we find $M_\nu = 0.24^{+0.34}_{-0.18}\ \mathrm{eV}$ and $\eta^2 = 2.14^{+0.30}_{-0.32}$ in the CMB framework, and $M_\nu = 0.37^{+0.34}_{-0.26}\ \mathrm{eV}$ and $\eta^2 = 2.4^{+2.1}_{-1.6}$ in the local framework. In particular, we find a 7$\sigma$ measurement of a non-zero neutrino asymmetry in the CMB framework. These neutrino parameters are consistent with those, in our previous work, obtained from the \emph{Planck} CMB temperature power spectrum. These results demonstrate that galaxy pairwise velocities provide an independent and sensitive probe of neutrino properties, opening a new avenue for testing neutrino physics with large-scale structure observations.
\end{abstract}

\maketitle

\section{Introduction}
Neutrinos are one of the fundamental fermions in the Universe, yet many of their properties are not fully understood. Neutrino oscillation experiments have established that at least two of the three neutrino mass eigenstates are massive, with masses $m_i$, $i = 1, 2, 3$, implying a lower bound on the sum of neutrino masses, $M_\nu \equiv \sum_i m_i$, of $M_\nu \gtrsim 0.06~\mathrm{eV}$ ($0.10~\mathrm{eV}$) for the normal (inverted) mass hierarchy~\citep{esteban2020fate}. In addition to upcoming laboratory experiments, such as the Karlsruhe Tritium Neutrino (KATRIN) experiment~\citep{Katrin2022NatPh,Katrin2025Sci} and Project~8~\citep{esfahani2022project,Esfahani2023PhRvL}, cosmology provides a powerful and independent, albeit model dependent, means to constrain neutrino masses.

Cosmic microwave background (CMB) observations, including those from \emph{Planck}~\citep{planck2020aa}, the Atacama Cosmology Telescope (ACT)~\cite{ACT2025JCAP}, and the South Pole Telescope (SPT-3G)~\cite{SPT3G2025arXiv}, currently provide some of the tightest upper limits on $M_\nu$. For instance, the combined analysis of \emph{Planck}, SPT-3G, and ACT CMB data (CMB-SPA) gives $M_\nu < 0.17~\mathrm{eV}$ (95\% CL)~\cite{SPT3G2025arXiv}.

More recently, the Dark Energy Spectroscopic Instrument (DESI DR2) reported a neutrino mass tension within the Lambda Cold Dark Matter ($\Lambda$CDM) framework when allowing $M_\nu$ to vary and jointly fitting CMB (\emph{Planck} and ACT), DESI DR2 baryon acoustic oscillation (BAO), and Type~Ia supernova data~\citep{DESIDR22025PRD}, yielding $M_\nu < 0.053~\mathrm{eV}$ (95\% CL). Large-scale structure (LSS) observations also provide complementary constraints on $M_\nu$. For example, recent weak-lensing measurements from the Kilo-Degree Survey (KiDS)~\cite{KiDS2025arxiv} report an upper bound of $M_\nu \lesssim 1.5~\mathrm{eV}$ (68\% CL).

Besides finite masses, relic neutrinos may also have a finite chemical potential, which is another important cosmological parameter. It is usually expressed through the dimensionless degeneracy parameter $\xi \equiv \mu/T_\nu$, where $\mu$ and $T_\nu$ are the neutrino chemical potential and temperature, respectively. Big Bang Nucleosynthesis (BBN) data tightly constrain the electron-type neutrino degeneracy parameter, $\xi_e$, as it directly affects the weak interaction rates that govern the neutron-to-proton ratio, requiring $\xi_e \ll 1$~\citep{Pitrou2018PhR,xinu_constrain2023_s1, xinu_constrain2023_s2}. In contrast, the muon- and tau-type neutrino degeneracy parameters are only indirectly constrained through their contribution to the total radiation energy density, commonly parameterized by the effective number of relativistic species, $\Delta N_\mathrm{eff}$, during the BBN epoch,
\begin{equation}
    \Delta N_\mathrm{eff} = \sum_i \left[\frac{30}{7}\left(\frac{\xi_i}{\pi}\right)^2 + \frac{15}{7}\left(\frac{\xi_i}{\pi}\right)^4\right],
\end{equation}
where $i = 1, 2, 3$ labels the neutrino mass eigenstates. The flavor and mass eigenstates are related by the Pontecorvo--Maki--Nakagawa--Sakata (PMNS) matrix (e.g., see Eq.~(3) of~\citet{barenboim2017LmuLtau}). Moreover, current CMB observations, which constrain $\xi_i$ directly through their impact on the perturbation evolution, still allow degeneracy parameters of order unity, $\xi_i \sim \mathcal{O}(1)$~\citep{yeung2021relic,Yeung2025ApJL,Zhao2025PRD}. 
Given the tight bounds on $\xi_e$ and the strong correlation between the muon and tau flavors, we assume $\xi_e = 0$ and $\xi_\mu = \xi_\tau$. Under this assumption, the neutrino degeneracy can be characterized by a single parameter, the asymmetry parameter $\eta^2\equiv\sum_i\xi_i^2$.~\footnote{Some studies use the third mass eigenstate degeneracy parameter $\xi_3$~\citep{yeung2021relic, Yeung2025ApJL} instead, which can be derived from $\eta^2$ simply. We therefore also present results in terms of $\xi_3$ to facilitate direct comparison with the literature.}  
The degeneracy between the total neutrino mass $M_\nu$ and the asymmetry parameter $\eta^2$ has been investigated using a variety of cosmological probes, including CMB observations~\citep{yeung2021relic, Yeung2025ApJL, Zhao2025PRD}, the matter power spectrum~\citep{Carton2019JCAP}, the halo mass function and bias~\citep{Liu2026}, as well as halo merger history~\citep{Wong2021JCAP}.

In this Letter, we focus on the mean pairwise peculiar velocity, a well-established cosmological observable that has been used to constrain $\sigma_8$ and $\Omega_m$~\citep{juszkiewicz2000evidence, feldman2003estimate, ma2015constraining,Garc2025arXiv}, the local growth rate $f\sigma_8$~\citep{Howlett2017MNRAS, dupuy2019estimation}, dark energy and modified gravity models~\citep{bhattacharya2008dark, bhattacharya2011galaxy, mueller2015constraints1, bibiano2017pairwise, jaber2024PRD, Luo2025arXiv}, as well as sterile neutrinos~\citep{Hu2025JCAP}.

More specifically,~\citet{Zhang2024MNRAS} studied the impact of neutrino mass and asymmetry on the pairwise velocity using numerical simulations, demonstrating the strong sensitivity of this statistics to both effects. Subsequently,~\citet{Zhang2025ApJL} developed and validated an observational pipeline that uses the galaxy pairwise velocity to measure the Hubble constant $H_0$ and the present-day matter density $\Omega_m$. In this Letter, we build upon these results by applying the established pairwise velocity pipeline to jointly constrain $M_\nu$ and $\eta^2$.

The remainder of this Letter is organized as follows. In Section~\ref{sc:measure_neutrino}, we describe our simulation setup and the pairwise velocity. The results are presented in Section~\ref{sc:results}, and we summarize and discuss our findings in Section~\ref{sc:summary_measure}.

\section{Methodology}\label{sc:measure_neutrino}
In this section, we describe the numerical simulations employed in this Letter and define the pairwise velocity used to constrain neutrino properties.

\subsection{Simulation settings}
We assume that relic neutrinos retain a frozen Fermi--Dirac distribution in momentum after decoupling from the primordial thermal bath. Consequently, the homogeneous background neutrino energy density in the presence of a chemical potential is
\begin{multline}\label{eq:neutrino_density}
    \rho_\nu(T_\nu) = \\
    \frac{1}{2\pi^2}\sum_{i=1}^3 \int_0^\infty 
    \left[
    \frac{E_i}{e^{p/T_\nu-\xi_i}+1}
    + \frac{E_i}{e^{p/T_\nu+\xi_i}+1}
    \right] p^2\,\mathrm{d}p,
\end{multline}
where the sum runs over the three neutrino mass eigenstates, $E_i^2=p^2+m_i^2$, and the two terms in the brackets correspond to neutrinos and antineutrinos, respectively. Throughout this Letter, we consider three degenerate neutrino masses, $m_1 = m_2 = m_3 = M_\nu/3$.

We run cosmological dark-matter-only (DMO) $N$-body simulations that include both $M_\nu$ and $\eta^2$, using a grid-based linear response method implemented in \texttt{Gadget-2}~\citep{Springel2005MNRAS,Ali2013MNRAS,Carton2019JCAP}. In this approach, the homogeneous neutrino background is given by the energy density in Eq.~(\ref{eq:neutrino_density}), while neutrino perturbations are evolved by solving the Vlasov equation on spatial grids and coupled self-consistently to the nonlinear evolution of CDM. Further technical details of the method can be found in~\citet{Carton2019JCAP} and~\citet{Zhang2024MNRAS}. 

Initial conditions of these simulations are generated at $z=99$ using \texttt{2LPTic}~\citep{crocce2006transients}, adopting an expansion history that consistently incorporates neutrino effects. The linear matter power spectra are computed with a modified version of \texttt{CAMB}~\citep{Lewis:1999bs}, which is extended to include the effects of $\eta^2$~\citep{yeung2021relic}.

A further nontrivial ingredient of the simulation setup is the choice of the background cosmology. Several cosmological parameters are currently subject to observational tensions (see, e.g.,~\citet{CosmoVerse2025PDU}), most notably the Hubble constant $H_0$, the value of which inferred from CMB data is in significant disagreement with that obtained from local distance-ladder measurements. Moreover, neutrino parameters are known to be degenerate with $H_0$ in CMB analyses~\citep{Yeung2025ApJL}. 

To account for this uncertainty, we adopt two distinct frameworks in our simulations, based on cosmological parameters derived from CMB and local distance ladder measurements, respectively. Within each framework, we explore a two-dimensional parameter grid in $M_\nu$ and $\eta^2$, with the $\Lambda$CDM case corresponding to the point $(M_\nu, \eta^2) = (0, 0)$. 
The grid ranges are chosen to fully cover the  parameter region explored in the Markov Chain Monte Carlo (MCMC) analysis within each framework, so that the interpolation of simulation predictions is performed within the simulated parameter space rather than relying on extrapolation.

\noindent
\textbf{CMB framework.}
We consider
\begin{equation*}
\begin{aligned}
M_\nu &\in \{0.06,\,0.15,\,0.24,\,0.42,\,0.60,\,1.20\}\ \mathrm{eV},\\
\eta^2 &\in \{0,\,0.4,\,0.8,\,1.4,\,2.0,\,3.0\}.
\end{aligned}
\end{equation*}
For each $(M_\nu,\eta^2)$ pair, we refit the cosmological parameters using a modified version of \texttt{CosmoMC} that incorporates neutrino mass and asymmetry effects~\citep{yeung2021relic}. We employ \emph{Planck}~2018 plikHM\_TTTEEE temperature and polarization data, together with the BAO data set used in the \emph{Planck}~2018 analysis~\citep{planck2020aa}. The resulting mean values of cosmological parameters are summarized in Table~\ref{tab:CMB_framework_simu_table} of Appendix~\ref{sc:CMB_framework_table}.

\noindent
\textbf{Local framework.}
We take 
\begin{equation*}
\begin{aligned}
M_\nu &\in \{0.06,\,0.15,\,0.24,\,0.60,\,0.90,\,1.20\}\ \mathrm{eV},\\
\eta^2 &\in \{0,\,0.4,\,0.8,\,2.0,\,3.0,\,5.0,\,8.0\}.
\end{aligned}
\end{equation*}
In this framework, the background cosmology is fixed to values favored by local measurements. We adopt $H_0 = 73.6\ \mathrm{km\,s^{-1}\,Mpc^{-1}}$ and $\Omega_m = 0.334$, as inferred from the Pantheon+ Type~Ia supernova analysis~\citep{Pantheon2022}. The total matter density is decomposed as $\Omega_m = \Omega_{cb} + \Omega_\nu$, where $\Omega_\nu$ ($\Omega_{cb}$) denotes the neutrino (CDM + baryon) energy density. The baryon density is fixed to $\Omega_b h^2 = 0.02244$ from BBN constraints~\citep{BBN2020JCAP}, and the primordial scalar spectrum parameters are set to the \emph{Planck}~2018 values $n_s = 0.9652$ and $A_s = 2.0968 \times 10^{-9}$.

\vspace{4pt}
All simulations are performed in a cubic volume of side length $250\,h^{-1}\mathrm{Mpc}$, containing $N_p = 1024^3$ CDM particles. 
The CDM particle mass and the gravitational softening length are around $10^9\,h^{-1}M_\odot$ and $6.1\,h^{-1}\mathrm{kpc}$, respectively.
Dark matter halos are identified using the \texttt{ROCKSTAR} halo finder~\citep{behroozi2012rockstar}, and all distances are converted from $h^{-1}\mathrm{Mpc}$ to $\mathrm{Mpc}$. We restrict our analysis to distinct host halos, with halo masses defined using the virial threshold~\citep{bryan1998ApJ}, $\Delta_{\mathrm{vir}} = 102$, relative to the critical density $\rho_{\mathrm{crit}}$. Halo centres and velocities follow the default \texttt{ROCKSTAR} definitions: the halo centre is computed from the mean position of particles in the innermost subgroup, while the halo velocity is defined as the average particle velocity within the inner 10\% of the virial radius~\citep{behroozi2012rockstar}. Finally, to ensure consistency with the underlying cosmology, we modify the expansion rate used by \texttt{ROCKSTAR} to account for the neutrino contribution, in analogy with the treatment adopted in \texttt{2LPTic}.

To reduce cosmic variance, we run three independent realizations of the reference model $(M_\nu = 0.06~\mathrm{eV},\, \eta^2 = 0)$ in both frameworks. Each realization includes four variants: standard, paired, fixed, and paired-and-fixed, following the method of~\citet{paired_and_fixed_2016MNRAS} and~\citet{ paired_and_fixed_2020MNRAS}.

\subsection{Pairwise velocity}

From these simulations, we compute the halo--halo pairwise velocity by
\begin{equation}\label{eq:pwv-definition}
    v_\mathrm{hh}(r) =
    \left\langle 
    \left[\boldsymbol{v}_1(\boldsymbol{r}_1) - \boldsymbol{v}_2(\boldsymbol{r}_2)\right]
    \cdot \hat{r}
    \right\rangle ,
\end{equation}
where $\boldsymbol{r}_{1,2}\ (\boldsymbol{v}_{1,2})$ are the positions (peculiar velocities) of halo 1 and 2, respectively, $\boldsymbol{r} \equiv \boldsymbol{r}_1 - \boldsymbol{r}_2$ is their comoving separation vector, and $\hat{r} \equiv \boldsymbol{r}/|\boldsymbol{r}|$. The ensemble average $\langle \cdots \rangle$ is taken over all pairs with separation $r \equiv |\boldsymbol{r}|$. 

As shown in Figure~4 of~\citet{Zhang2024MNRAS}, the halo--halo pairwise velocity depends sensitively on the halo mass range. Following~\citet{Zhang2025ApJL}, we therefore select halos with masses in the range $[10^{11},\,10^{13}]\,M_\odot$, so as to match the halo masses of the Cosmicflows-4 (CF-4) grouped catalog~\citep{Tully2023ApJ} used in this Letter. The halo--halo pairwise velocity quantifies the mean relative motion of halo pairs as a function of the halo separation. A typical halo--halo pairwise velocity is shown in the top panels of Figure~\ref{fig:neutrino_effects_local_CMB}. Over the full range of separations considered, $v_{\mathrm{hh}}(r)$ is negative, indicating that halo pairs preferentially approach each other as a result of gravitational attraction. The magnitude of $v_{\mathrm{hh}}(r)$ reaches a minimum at $r \sim 7\,\mathrm{Mpc}$, corresponding to the characteristic clustering scale of halos in the mass range $[10^{11},\,10^{13}]\,M_\odot$, where this approaching tendency is strongest. These features are consistent with our previous analysis in~\citet{Zhang2024MNRAS}, where a more detailed discussion can be found.

The galaxy--galaxy pairwise velocity is estimated using galaxies from the CF-4 grouped catalog~\citep{Tully2023ApJ}. 
In observations, only line-of-sight velocities are accessible, inferred from galaxy distances and redshifts under an assumed background cosmology. As a result, the pairwise velocity cannot be measured directly and must be reconstructed statistically from line-of-sight peculiar velocities of galaxy pairs. Following~\citet{Ferreira1999APJL}, the galaxy--galaxy pairwise velocity is estimated by
\begin{equation}
    v_{\rm gg}(r) =
    \frac{\sum (v_{p,1} - v_{p,2})\, p_{12}}
         {\sum p_{12}^2},
\end{equation}
where $v_{p,1/2}$ are the inferred line-of-sight peculiar velocities of galaxies 1 and 2, respectively, and
$p_{12} \equiv \hat{r} \cdot (\hat{r}_1 + \hat{r}_2)/2$.
The line-of-sight peculiar velocities are inferred from the observed CMB-rest-frame velocities and luminosity distances under an assumed background cosmology, and therefore they depend on $H_0$ and $\Omega_m$. Details of the estimator, the inference of peculiar velocities, and the treatment of observational uncertainties are presented in~\citet{Zhang2025ApJL} and not repeated here.

Throughout this Letter, halo--halo pairwise velocities from simulations are evaluated at $z=0$. For both halo and galaxy pairwise velocities, we adopt an identical binning scheme, using eight linear separation bins over $r \in [0,\,16]~\mathrm{Mpc}$, consistent with our previous work~\citep{Zhang2025ApJL}.

\section{Results}\label{sc:results}

\begin{figure*}
    \centering
    \includegraphics[width=0.48\linewidth]{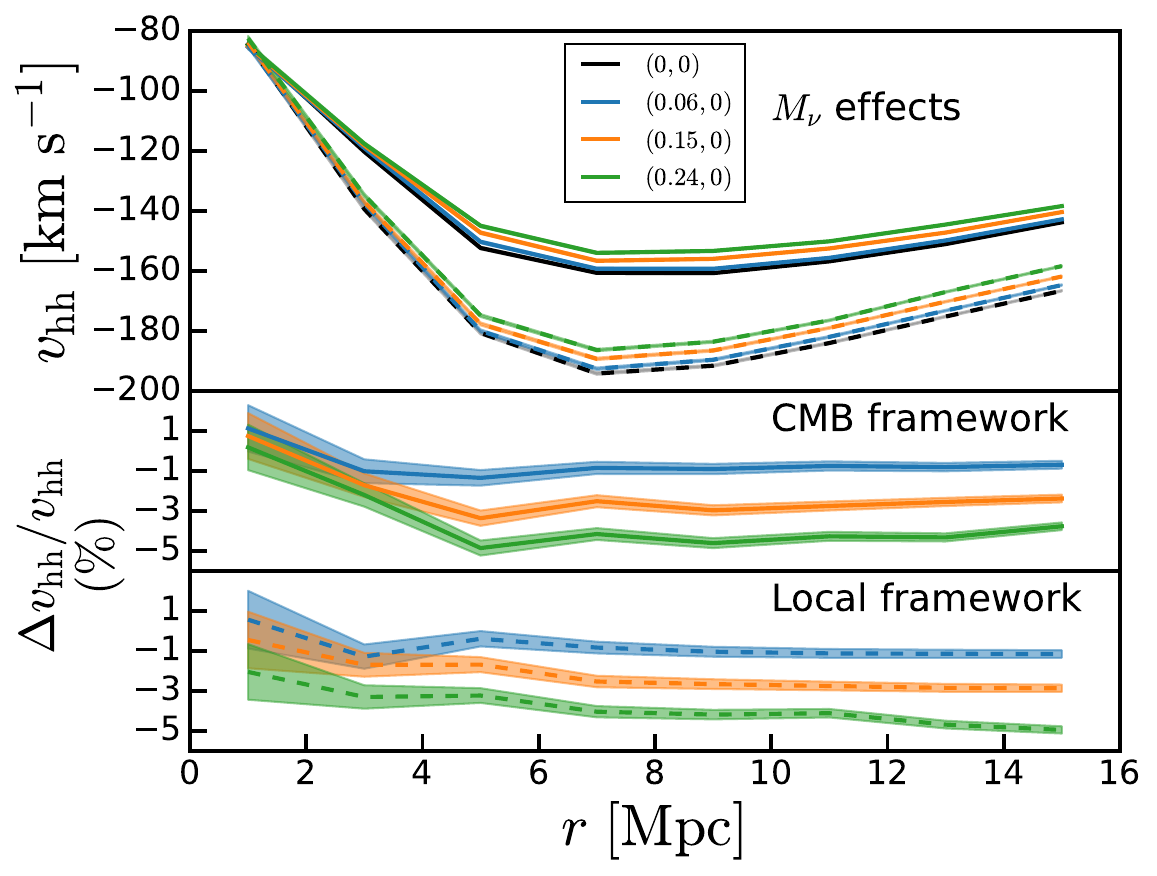}
    \includegraphics[width=0.48\linewidth]{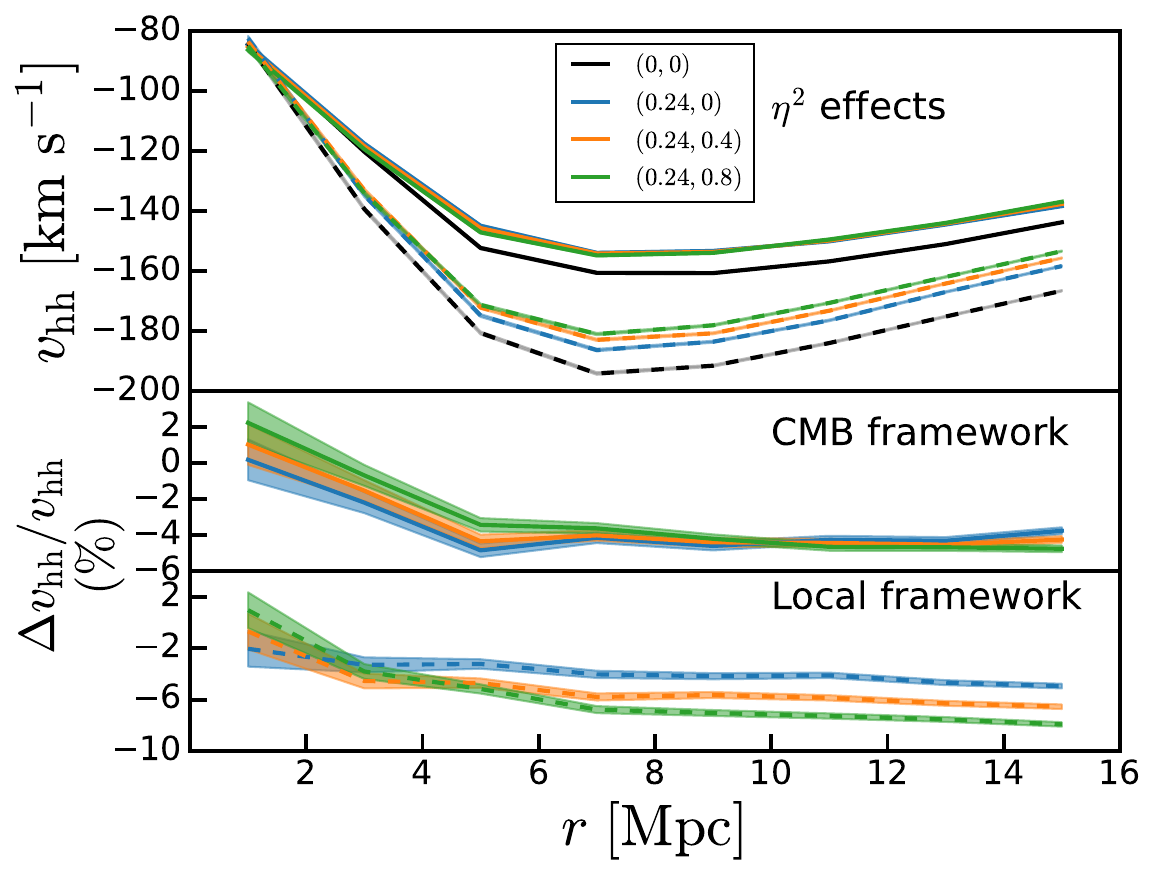}
    \caption{Neutrino effects on the halo--halo pairwise velocity $v_{\mathrm{hh}}$ in the CMB (solid lines) and local (dashed lines) frameworks. Left and right panels show the effects of varying $M_\nu$ and $\eta^2$, respectively. Top panels show $v_{\mathrm{hh}}(r)$ for various ($M_\nu$,$\eta^2$) combinations. The middle and bottom panels show $\Delta v_\mathrm{hh}/v_\mathrm{hh} = v_\mathrm{hh}(M_\nu, \eta^2)/v_\mathrm{hh} (0, 0) - 1$ in the CMB and local frameworks, respectively.}    
    \label{fig:neutrino_effects_local_CMB}
\end{figure*}

\begin{figure*}
    \centering
    \includegraphics[width=0.85\linewidth]{neutrino_density_plot.pdf}
    \caption{2D projected CDM density fields in the local framework at $z=0$, shown on different spatial scales. The left, middle, and right columns correspond to $(M_\nu,\eta^2)$ = (0, 0), (0.24, 0), and (0.24, 0.8), respectively. The top row shows projections over a volume of $80 \times 80 \times 20~\mathrm{Mpc}^3$, with white squares indicating the regions shown in the bottom row. The bottom row presents magnified views of the selected regions on scales of $10 \times 10 \times 10~\mathrm{Mpc}^3$. White circles mark the virial radius $R_{\rm vir}$ of the central halo, labeled together with the corresponding virial mass $M_{\rm vir}$ at the upper left of each panel. The density colour bars are shown on the right-hand side.
    }
    \label{fig:density_plot_local_framework}
\end{figure*}

Figure~\ref{fig:neutrino_effects_local_CMB} shows the effects of varying $M_\nu$ and $\eta^2$ on the halo--halo pairwise velocity $v_\mathrm{hh}$ in both the local and CMB frameworks.

In the local framework, the effects of $M_\nu$ and $\eta^2$ on the pairwise velocity arise from modifications of the background expansion history and neutrino free streaming.
We test the expansion-history contribution and find its impact on the pairwise velocity to be subdominant at the percent level, consistent with previous findings for the matter power spectrum (see, e.g., Figure~1 of~\citet{Carton2019JCAP}).
Both a nonzero $M_\nu$ and a nonzero $\eta^2$ increase the neutrino energy density (Eq.~\eqref{eq:neutrino_density}), 
thereby amplifying the suppression of structure growth caused by neutrino free streaming.
The resulting suppression of structure formation leads to a qualitatively similar response of the pairwise velocity to $M_\nu$ and $\eta^2$, characterized by an approximately scale-independent reduction across the separations considered. This interpretation is further supported by the projected density fields in Figure~\ref{fig:density_plot_local_framework}, where increasing $\eta^2$ produces a smoother large-scale structure, similar to the effect of increasing $M_\nu$.

However, the behavior is qualitatively different in the CMB framework.
In addition to the effects of the expansion history and neutrino free streaming, each $(M_\nu,\eta^2)$ case in the CMB framework is associated with a different set of refitted cosmological parameters.
As a result, increasing $M_\nu$ produces an approximately scale independent suppression of $v_{\rm hh}$, while a nonzero $\eta^2$ leads to a scale dependent response, with $v_{\rm hh}$ enhanced on small scales but suppressed on large scales.
This behavior is consistent with the results of~\citet{Zhang2024MNRAS}, where similar trends are observed in the CMB framework, including in the projected density fields shown in their Figure~1.

To quantify neutrino effects, we adopt a linear regression approach, following the similar methodology in~\citet{Zhang2025ApJL}. For both the CMB and local frameworks, the halo--halo pairwise velocity is modeled as
\begin{multline}\label{eq:fit_pwv_Mnu_eta}
    v_\mathrm{hh}(M_\nu,\eta^2,r) = \bar{v}_\mathrm{hh}(0.06,0,r) \\
    \times \left[1 + C_m(r)\frac{M_\nu - 0.06\ \mathrm{eV}}{0.06\ \mathrm{eV}} + C_\eta(r)\eta^2\right],
\end{multline}
where $C_m(r)$ and $C_\eta(r)$ are scale-dependent fitting coefficients, and $\bar{v}_\mathrm{hh}(0.06,0,r)$ denotes the fiducial pairwise velocity averaged over 12 simulations to suppress cosmic variance. The regression is performed at a fixed random seed to obtain the fractional variation induced by neutrino parameters. Although the same fitting form is employed across both frameworks, the fitting coefficients $C_m(r)$, $C_\eta(r)$, and $\bar{v}_\mathrm{hh}(0.06,0,r)$ are determined independently for each framework.

Since the galaxy--galaxy pairwise velocity $v_{\mathrm{gg}}$ depends on both $H_0$ and $\Omega_m$ as shown in~\citet{Zhang2025ApJL}, we adopt different treatments for these parameters in the two frameworks. In the local framework, we fix $H_0 = 73.6~\mathrm{km\,s^{-1}\,Mpc^{-1}}$ and $\Omega_m = 0.334$. In the CMB framework, $H_0$ and $\Omega_m$ are treated as derived quantities that vary with neutrino parameters $(M_\nu, \eta^2)$, parameterized as
\begin{equation}\label{eq:fit_H0_Om_Mnu_eta}
    \begin{aligned}
        H_0(M_\nu,\eta^2) &= H_0(0.06,0)\left[1 + C_m^{H}\frac{M_\nu - 0.06\ \mathrm{eV}}{0.06\ \mathrm{eV}} + C_\eta^{H}\eta^2\right], \\
        \Omega_m(M_\nu,\eta^2) &= \Omega_m(0.06,0)\left[1 + C_m^{\Omega}\frac{M_\nu - 0.06\ \mathrm{eV}}{0.06\ \mathrm{eV}} + C_\eta^{\Omega}\eta^2\right],
    \end{aligned}
\end{equation}
where $H_0(0.06,0)$ and $\Omega_m(0.06,0)$ denote the reference values corresponding to $M_\nu = 0.06~\mathrm{eV}$ and $\eta^2 = 0$. By fitting the parameters in Table~\ref{tab:CMB_framework_simu_table}, we obtain the best-fit coefficients as
\begin{equation}
    \begin{aligned}
        C_m^{H} &= -0.005649 \pm 0.000089, 
         & C_\eta^{H} &= 0.04027 \pm 0.00050, \\
        C_m^{\Omega} &= 0.014452 \pm 0.000063, 
         & C_\eta^{\Omega} &= -0.02543 \pm 0.00035.
    \end{aligned}
\end{equation}

Following the methodology in~\citet{Zhang2025ApJL}, we define the $\chi^2$ as
\begin{equation}
    \chi^2(M_\nu,\eta^2) = \sum_{ij} D_v(M_\nu,\eta^2,r_i) C^{-1}_{ij} D_v(M_\nu,\eta^2,r_j),
\end{equation}
where $D_v(r) \equiv v_{\mathrm{gg}}(r) - v_{\mathrm{hh}}(r)$, and the covariance matrix $C_{ij} = [C_{\mathrm{gg}}]_{ij} + (\Delta v_{\mathrm{hh}})^2 \delta_{ij}$, with $\Delta v_{\mathrm{hh}}$ representing the fitting uncertainties from Eq.~\eqref{eq:fit_pwv_Mnu_eta}. The observational covariance $C_{\mathrm{gg}}$ is described in~\citet{Zhang2025ApJL}. The likelihood at each MCMC step is computed as
\begin{equation}\label{eq:likelihood_Mnu_eta2}
    \mathcal{L}(M_\nu,\eta^2) \sim e^{-\chi^2(M_\nu,\eta^2)/2}P(M_\nu)P(\eta^2),
\end{equation}
where $P(M_\nu)$ and $P(\eta^2)$ are the prior distributions. 

Thus, we perform MCMC sampling using \texttt{emcee}~\citep{emcee2013}, adopting flat priors $M_\nu \in [0, 2]~\mathrm{eV}$ and $\eta^2 \in [0, 10]$. 
The observational processing of the CF-4 catalog follows the procedure described in~\citet{Zhang2025ApJL}, and those details are not repeated here.

At each MCMC step, given $(M_\nu, \eta^2)$, $v_{\mathrm{hh}}(r)$ is computed using Eq.~\eqref{eq:fit_pwv_Mnu_eta}. In the CMB framework, the corresponding cosmological parameters $(H_0, \Omega_m)$ are derived from $(M_\nu, \eta^2)$ using Eq.~\eqref{eq:fit_H0_Om_Mnu_eta} and propagated through the observational pipeline described in~\citet{Zhang2025ApJL} to obtain the galaxy--galaxy pairwise velocity $v_{\mathrm{gg}}(r)$. In the local framework, $v_{\mathrm{gg}}(r)$ is obtained by fixing the cosmological parameters to $H_0 = 73.6~\mathrm{km\,s^{-1}\,Mpc^{-1}}$ and $\Omega_m = 0.334$. For both frameworks, the likelihood is evaluated by comparing $v_{\mathrm{hh}}(r)$ and $v_{\mathrm{gg}}(r)$ according to Eq.~\eqref{eq:likelihood_Mnu_eta2}.

\begin{figure*}
    \centering
    \includegraphics[width=0.48\linewidth]{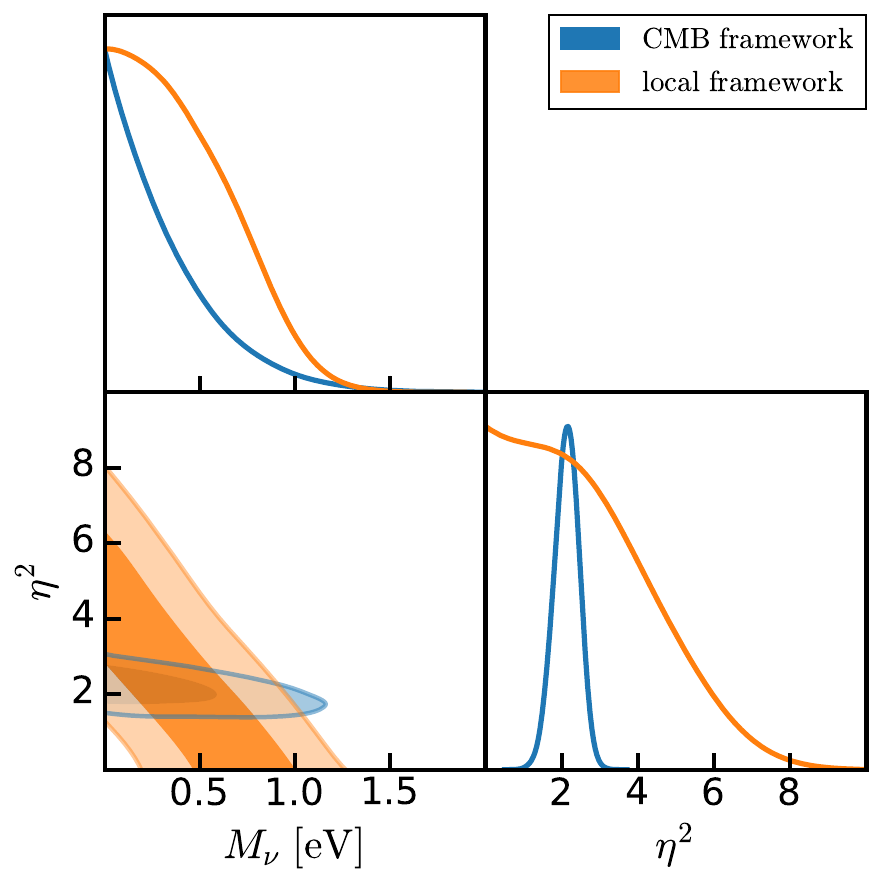}
    \includegraphics[width=0.5\linewidth]{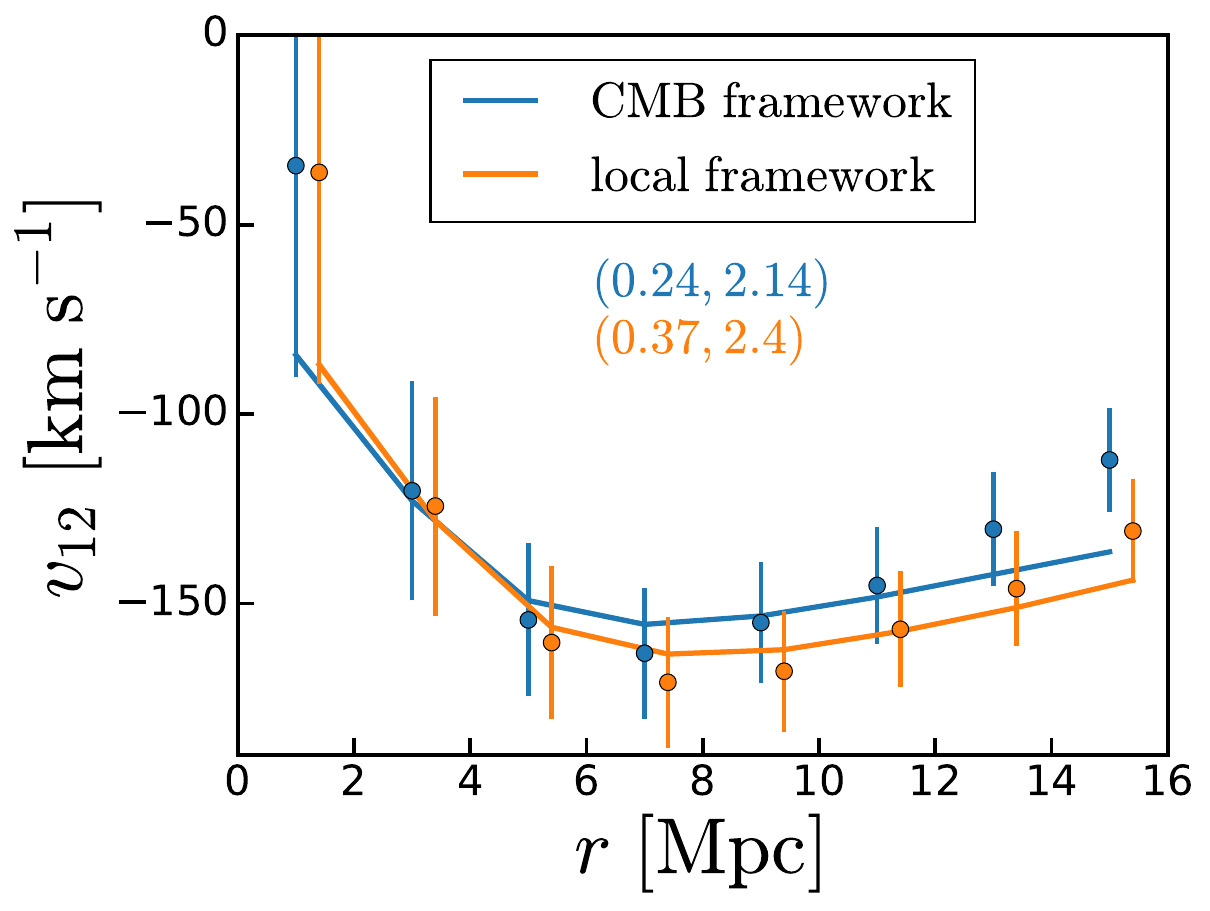}
    \caption{Left: Marginalized posterior distributions and 68\%/95\% confidence contours for $(M_\nu, \eta^2)$ in the CMB (blue lines) and local (orange lines) frameworks. Right: Comparison between $v_{\mathrm{hh}}$ (solid lines, from simulations) and $v_{\mathrm{gg}}$ (points with error bars, from observations), evaluated at the corresponding mean posterior values in the two frameworks of $(M_\nu, \eta^2)$, labelled in the corresponding colours. For clarity, the local framework data is shifted horizontally slightly.}
    \label{fig:fit_local_CMB_best_fit}
\end{figure*}

\begin{table}
    \centering
    \begin{tabular}{cccc}
        \hline
        \hline
         &  $M_\nu\ [\mathrm{eV}]$& $\eta^2$ &$\xi_3$\\
        \hline
         CMB framework&  $0.24^{+0.34}_{-0.18}$& $2.14^{+0.30}_{-0.32}$&$1.286^{+0.082}_{-0.096}$\\
         Local framework&  $0.37^{+0.34}_{-0.26}$& $2.4^{+2.1}_{-1.6}$ &$1.35^{+0.47}_{-0.58}$\\
        \hline
        \hline        
    \end{tabular}
    \caption{Posterior means and 68\% CL for $M_\nu$, $\eta^2$, and the derived parameter $\xi_3$, obtained from both the CMB and local frameworks.}
    \label{tab:Mnu_eta2_bestfit}
\end{table}

The inferred constraints on $M_\nu$ and $\eta^2$ are summarized in Table~\ref{tab:Mnu_eta2_bestfit} and are quoted as posterior means with 68\% CL. In the CMB framework, we obtain $M_\nu = 0.24^{+0.34}_{-0.18}\,\mathrm{eV}$ and $\eta^2 = 2.14^{+0.30}_{-0.32}$, while in the local framework we find $M_\nu = 0.37^{+0.34}_{-0.26}\,\mathrm{eV}$ and $\eta^2 = 2.4^{+2.1}_{-1.6}$. The marginalized posterior distributions, together with comparisons between the halo--halo and galaxy--galaxy pairwise velocities ($v_{\rm hh}$ and $v_{\rm gg}$) evaluated at the mean values of $M_\nu$ and $\eta^2$, are shown in Figure~\ref{fig:fit_local_CMB_best_fit}.

In the local framework, $M_\nu$ and $\eta^2$ exhibit a clear anti-correlation. As illustrated in Figure~\ref{fig:neutrino_effects_local_CMB}, the effects of increasing $M_\nu$ and $\eta^2$ on $v_{\rm hh}$ are largely degenerate, leading to similar suppression of the pairwise velocity magnitude. In contrast, in the CMB framework, while increasing $M_\nu$ leads to an approximately scale-independent suppression of $v_{\rm hh}$, the effect of a nonzero $\eta^2$ depends on $r$. Moreover, in the CMB framework, each $(M_\nu,\eta^2)$ combination is accompanied by a self-consistent refitting of other cosmological parameters constrained by CMB data, such as $H_0$ and $\Omega_m$, which further modifies the galaxy--galaxy pairwise velocity. Therefore, there is less degeneracy between $M_\nu$ and $\eta^2$, resulting in tighter constraints on them compared to the local framework.

In both the CMB and local frameworks, the galaxy pairwise velocity data prefers a nonzero neutrino asymmetry, reaching a 7$\sigma$ preference in the former. For completeness, we also report in Table~\ref{tab:Mnu_eta2_bestfit} the corresponding values of the degeneracy parameter of the third neutrino mass eigenstate, $\xi_3$, inferred in both frameworks. Notably, these inferred values are consistent with the constraint from CMB temperature data reported in~\citet{Yeung2025ApJL}, $\xi_3 = 1.13^{+0.41}_{-0.19}$.

\section{Conclusion}\label{sc:summary_measure}

In this Letter, we present the first measurements of the total neutrino mass $M_\nu$ and asymmetry parameter $\eta^2$ derived from the mean galaxy pairwise peculiar velocity in the quasi-linear and nonlinear regimes. Our analysis is carried out within two independent frameworks, based on cosmological parameters derived from CMB and local distance ladder measurements, respectively. By performing MCMC fits of simulation-based models to galaxy data from the CF-4 grouped catalog, we obtain the following posterior means with 68\% CL:
\begin{itemize}
    \item[(i)] In the CMB framework, $M_\nu = 0.24^{+0.34}_{-0.18}\ \mathrm{eV}$ and $\eta^2=2.14^{+0.30}_{-0.32}$.
    \item[(ii)] In the local framework, $M_\nu = 0.37^{+0.34}_{-0.26}\ \mathrm{eV}$ and $\eta^2=2.4^{+2.1}_{-1.6}$.
\end{itemize}

The constraints derived from the two frameworks are consistent with each other within uncertainties. Notably, in the CMB framework analysis, we find a 7$\sigma$ preference for a nonzero $\eta^2$. Furthermore, the inferred values of $\xi_3$ from both frameworks are consistent with that obtained from CMB temperature data, as reported in~\citet{Yeung2025ApJL}. These results demonstrate the potential of the galaxy pairwise velocity as an independent and sensitive probe of neutrino properties in cosmology.

\vspace{0.4em}

\begin{acknowledgments}
The computational resources used for the simulations in this Letter were provided by the Central Research Computing Cluster at the Chinese University of Hong Kong. This research is supported by grants from the Research Grants Council of the Hong Kong Special Administrative Region, China, under Project Nos.\ AoE/P-404/18 and 14300223. WZ acknowledges support from the French Agence Nationale de la Recherche (ANR), under grant ANR-23-CE31-0010 (project ProGraceRay). SL acknowledges the supports by the National Natural Science Foundation of China (NSFC) grant (Nos 12588202, 12473015). All figures in this Letter were generated using \texttt{Matplotlib}~\citep{matlibplot2007} and \texttt{GetDist}~\citep{lewis2025JCAP}, together with \texttt{SciPy}~\citep{scipy2020} and \texttt{NumPy}~\citep{numpy2020}.
\end{acknowledgments}

\appendix
\section{CMB framework cosmological parameters}\label{sc:CMB_framework_table}

Table~\ref{tab:CMB_framework_simu_table} lists the refitted cosmological parameters used in our $N$-body simulations.

\begin{table*}
    \centering
    \begin{tabular}{ccccccc}
         \hline\hline    
         $M_\nu$ [eV] & $\eta^2$ & $H_0$ [km s$^{-1}$ Mpc$^{-1}$] & $\Omega_m$ & $\Omega_bh^2$& $n_s$ & $A_s\ [10^{-9}]$ \\
         \hline    
         0.00 & 0.0 & 68.03 & 0.3070 & 0.02241 & 0.9661 & 2.100 \\
         0.06 & 0.0 & 67.73 & 0.3105 & 0.02242 & 0.9667 & 2.102 \\
         0.06 & 0.4 & 68.77 & 0.3075 & 0.02255 & 0.9728 & 2.120 \\
         0.06 & 0.8 & 69.88 & 0.3043 & 0.02268 & 0.9789 & 2.138 \\
 0.06& 1.4& 71.60&  0.2994&   0.02288&  0.9883&2.167\\
 0.06& 2.0& 73.40&  0.2943&  0.02307&  0.9978& 2.192\\
 0.06& 3.0&  76.60&  0.2855&  0.02339&  1.0142& 2.235\\
         0.15 & 0.0 & 67.14 & 0.3169 & 0.02244 & 0.9680 & 2.106 \\
         0.15 & 0.4 & 68.18 & 0.3138 & 0.02257 & 0.9741 & 2.123 \\
         0.15 & 0.8 & 69.26 & 0.3107 & 0.02270 & 0.9802 & 2.143 \\
 0.15& 1.4& 70.92&  0.3061&  0.02290&   0.9894&  2.169\\
         0.15 & 2.0 &  72.69&  0.3012&  0.02310&  0.9990&  2.197\\
         0.15 & 3.0 &  75.82&  0.2925&  0.02342&    1.0155&   2.243\\
         0.24 & 0.0 & 66.55 & 0.3234 & 0.02246 & 0.9691 & 2.107 \\
         0.24 & 0.4 & 67.58 & 0.3202 & 0.02260 & 0.9752 & 2.127 \\
         0.24 & 0.8 & 68.62 & 0.3174 & 0.02273 & 0.9814 & 2.146 \\
 0.24& 1.4&  70.25&  0.3129&  0.02292&  0.9908&2.175\\
 0.24& 2.0&  72.00&  0.3079&  0.02312&  1.0005&  2.203\\
 0.24& 3.0&  75.06&  0.2994&   0.02345&  1.0170& 2.248\\
 0.42& 0.0&  65.42&  0.3364&  0.02250&  0.9713&2.113\\
 0.42& 0.4&  66.37&  0.3340&  0.02263&  0.9774& 2.132\\
 0.42& 0.8&  67.37&  0.3313&  0.02277&  0.9836& 2.154\\
 0.42& 1.4&  68.95&  0.3268&  0.02296&  0.9932& 2.181\\
 0.42& 2.0&  70.63&  0.3218&  0.02317&  1.0030& 2.212\\
 0.42& 3.0&  73.59&  0.3134&  0.02350&  1.0196& 2.261\\
 0.6& 0.0&  64.37&  0.3493& 0.02254&  0.9732& 2.116\\
 0.6& 0.4&  65.27&  0.3472&  0.02267&  0.9792& 2.136\\
 0.6& 0.8&  66.24&  0.3444&  0.02280&  0.9856& 2.155\\
 0.6& 1.4&  67.75&  0.3404&  0.02301&  0.9952&2.185\\
         0.6& 2.0 &  69.36&  0.3356&  0.02321&  1.0052&  2.218\\
         0.6& 3.0 &  72.25&  0.3270&  0.02355&  1.0219&  2.272\\
 1.2& 0.0&  61.23&  0.3930&  0.02265&  0.9765& 2.120\\
 1.2& 0.4&  62.03&  0.3914&  0.02279&  0.9829& 2.140\\
 1.2& 0.8&  62.86&  0.3897&  0.02294&  0.9893& 2.164\\
 1.2& 1.4&  64.21&  0.3862&  0.02316&  0.9990&  2.197\\
         1.2& 2.0 &  65.67&  0.3819&  0.02337&  1.0091&  2.232\\
         1.2& 3.0 &  68.32&  0.3733&  0.02375&  1.0263&  2.303\\
         \hline\hline
    \end{tabular}
    \caption{Refitted cosmological parameters in the CMB framework. The total matter density is the sum of CDM, baryon, and neutrino densities, i.e. $\Omega_m = \Omega_{cb} + \Omega_{\nu}$.}
    \label{tab:CMB_framework_simu_table}
\end{table*}

\end{document}